\documentclass[%
 reprint,
 amsmath,amssymb,
 aps,
 prl,
]{revtex4-2}

\usepackage{graphicx}
\usepackage{dcolumn}
\usepackage{bm}
\usepackage[dvipsnames]{xcolor}
\usepackage{subfig}

\begin{document}

\preprint{APS/123-QED}

\title{New plasma regimes with small ELMs and high confinement at the Joint European Torus}

\author{J.Garcia${^1}$, E. de la Luna${^2}$,  M. Sertoli${^3}$, F. J. Casson${^3}$, S. Mazzi$^{1,4}$, 
\v{Z}. \v{S}tancar$^{5}$, G. Szepesi$^{3}$, D. Frigione${^6}$, L. Garzotti${^3}$, F. Rimini${^3}$, D. van Eester${^7}$, P. Lomas${^3}$, C. Sozzi${^8}$, R. N. Aiba$^{9}$, R. Coelho${^{10}}$, L. Frasinetti${^{11}}$, G. Huijsmans${^{1,12}}$, F. Liu${^1}$  and JET contributors$^{a}$}
\affiliation{
${^1}$CEA, IRFM, F-13108 Saint Paul-lez-Durance, France.
\\${^2}$Laboratorio Nacional de Fusión, CIEMAT, 28040 Madrid, Spain
\\${^3}$Culham Centre for Fusion Energy of UKAEA, Culham Science Centre, Abingdon, United Kingdom
\\${^4}$Aix-Marseille Université, CNRS PIIM, UMR 7345 Marseille, France
\\${^5}$Jožef Stefan Institute, Jamova cesta 39, SI-1000 Ljubljana, Slovenia
\\${^6}$National Agency for New Technologies, Energy and Sustainable Economic Development, ENEA, C.R. Frascati, Roma, Italy
\\${^7}$LPP-ERM/KMS, EUROfusion Consortium Member - Trilateral Euregio Cluster, Brussels, Belgium
\\${^8}$Istituto per la Scienza e Tecnologia dei Plasmi, 20125 Milano, Italy
\\${^9}$National Institutes for Quantum and Radiological Science and Technology, Naka, Ibaraki 311-0193, Japan
\\${^{10}}$Instituto de Plasmas e Fusão Nuclear, Instituto Superior Técnico, Universidade de Lisboa, 1049-001 Lisboa, Portugal
\\${^{11}}$Division of Fusion Plasma Physics, KTH Royal Institute of Technology, Stockholm, Sweden
\\${^{12}}$Eindhoven University of Technology, P.O. Box 513, 5600 MB Eindhoven, The Netherlands
\\$^a$See the author list of E. Joffrin et al., Nucl. Fusion 59, 112021 (2019)
}%

\date{\today}

\begin{abstract}

New plasma regimes with high confinement, low core impurity accumulation and small Edge-localized mode (ELMs) perturbations have been obtained close to ITER conditions in magnetically confined plasmas from the Joint European torus (JET) tokamak. Such regimes are achieved by means of optimized particle fuelling conditions which trigger a self-organize state with a strong increase in rotation and ion temperature and a decrease of the edge density. An interplay between core and edge plasma regions leads to reduced turbulence levels and outward impurity convection. These results pave the way to an attractive alternative to the standard plasmas considered for fusion energy generation in a tokamak with metallic wall environment such as the ones expected in ITER. 

\end{abstract}

\maketitle


\emph{Introduction}.$-$The construction of the tokamak ITER \cite{shimada2007progress} will clarify the possibilities of nuclear fusion of deuterium (D) and Tritium (T) by magnetically confined plasmas as a reliable source of energy. ITER operation will combine a significant number of challenges such as the simultaneously requirements of high stability and confinement of plasmas. From present day experiments, it is well known that confined plasmas lead to instabilities of different nature, such as microturbulence or MagnetoHydroDynamics (MHD) which are generated by strong temperature and density gradients and that can severely degrade the thermal confinement. Furthermore, the presence of a metal wall, necessary to avoid tritium retention, generates high Z impurities that can be transported to the plasma core by such gradients, polluting the plasma and generating unacceptable levels of radiated power which can damage the plasma confinement.

The preferred mode of operation in ITER, the H-mode \cite{wagnerhmode}, exemplifies such difficulties. Whereas the confinement is enhanced with respect other operation regimes due to the build-up of the so called pedestal at the plasma edge, i.e. a region where heat and particle transport is almost suppressed, the strong gradient formed leads to MHD instabilities called Edge Localized Modes (ELMs) \cite{Zohm_1996}. ELMs result in periodic, rapid expulsion of edge plasma leading to strong heat and particle fluxes into the plasma facing components which can be severely damaged. Simultaneously, the experience in tokamaks with metal walls show that ELMs are beneficial for preventing impurity penetration in the plasma core as they regularly flush the generated tungsten (W) from the plasma facing components when their frequency is high enough \cite{Gruber_NF2009} \cite{Neu_PoP2013}. However, high frequency ELMs, usually obtained in conditions of strong particle fuelling by gas puff, lead to thermal energy confinement degradation \cite{Neu_PoP2013}. On the other hand, good thermal core energy confinement, usually obtained in low frequency ELMs conditions, inherently lead to core plasma contamination with high Z impurities in the presence of a metal wall \cite{Angioni_2014}. Therefore, finding a plasma state with high thermal confinement in the presence of ELMs and a metal wall has become a challenge in the magnetically confined fusion field which poses difficulties to the general concept of energy generation by magnetically confined plasmas. 

Several techniques have been put forward to mitigate or even fully suppress ELMs while keeping good plasma properties. Among others, the Quiescent H-mode (QH-mode) \cite{burell_QH} or the use of active ELM control techniques such as externally applied magnetic perturbations \cite{Evans_RMP}. However, the applicability of such techniques to ITER has not been fully demonstrated yet in plasma conditions close to the ones expected in ITER.

Another alternative proposed is the use of ELMs regimes characterised by small pedestal perturbations, such as Type-III \cite{Zohm_1996}, grassy \cite{Kamada_2000} or Type-II ELMs \cite{Stober_2001,Stober_2005}, however such regimes require particular plasma conditions difficult to obtain in ITER \cite{Viezzer_2018} or lead to low thermal energy confinement \cite{Sartori_2004}.

\begin{figure*}[t!]
	\subfloat{\includegraphics[width=0.59\textwidth]{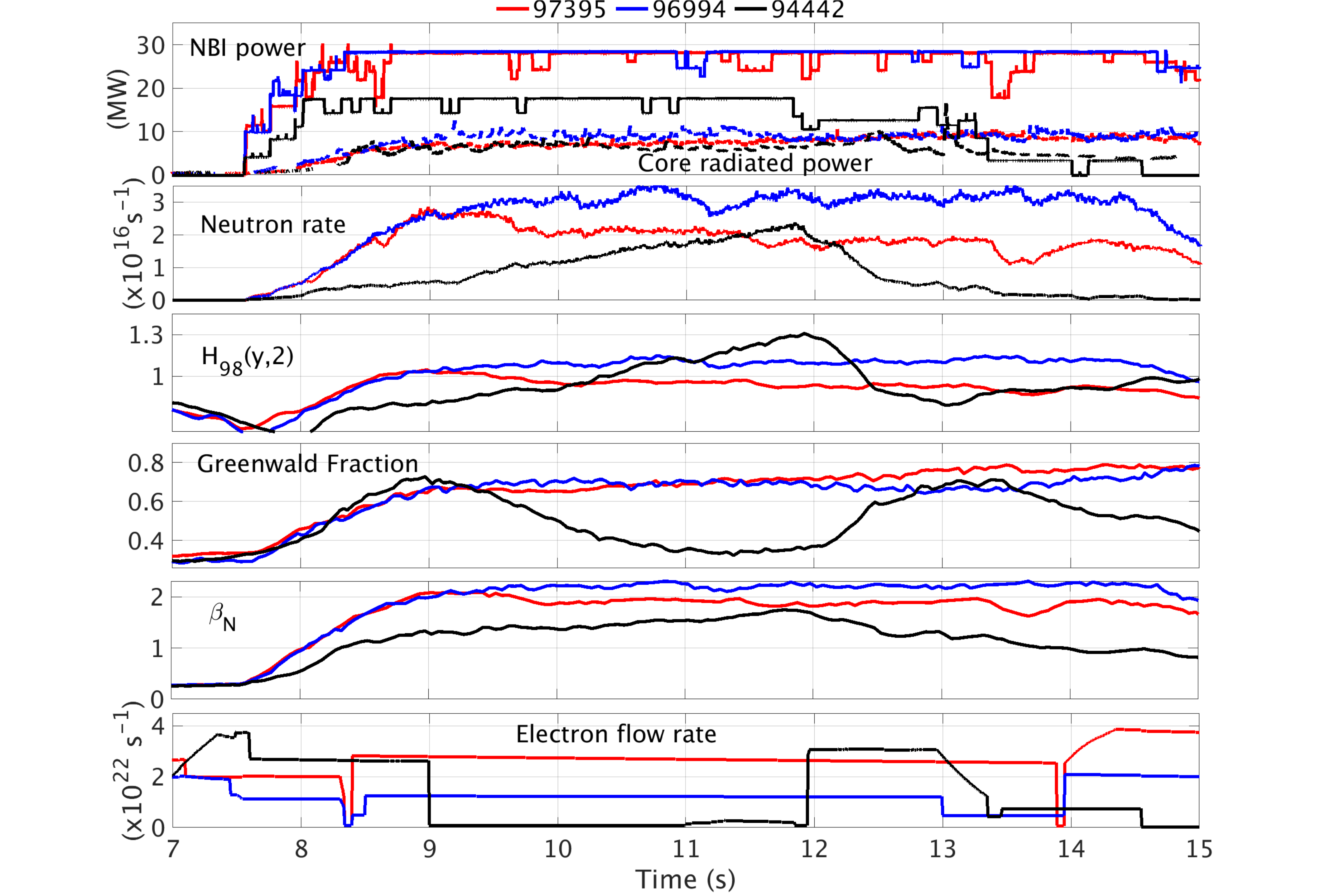}}
    \subfloat{\includegraphics[width=0.34\textwidth]{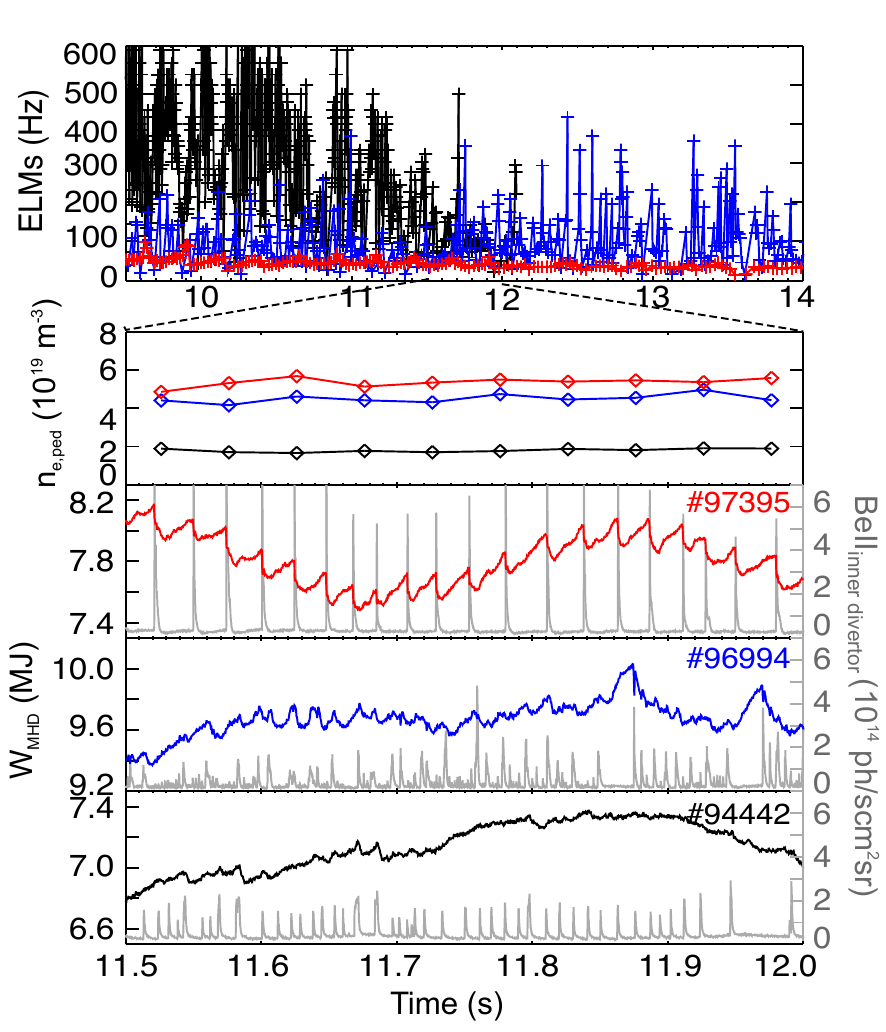}}
	\caption{Time evolution summary of the type I-ELM discharge \#97395 and small ELMs discharges \#94442 and \#96994. $H_{98}(y,2)$ is defined as the thermal energy with respect to the energy from the IPB98 scaling \cite{Transport_1999}. $\beta_N=\beta{a B_T}/{I_p} [\%]$ with $\beta$ the ratio between magnetic and thermal pressure (left). ELM frequency, pedestal density ($n_{e,ped}$), total plasma energy $W_{MHD}$ and BeII line emission from the inner divertor for the discharges \#97395, \#96994 and \#94442 (right).}
	\label{summary}
\end{figure*}

Until now, no adequate solution integrating good confinement, high neutron rate production, low heat fluxes to the plasma facing components and impurity core control has been found in plasmas close to ITER conditions, specially in the presence of a metallic wall. The results presented in this Letter show for the first time that small ELMs regimes are compatible with high core thermal energy confinement and fusion power at the Joint European Torus (JET)\cite{litaudon2017overview}, which is the existing closest tokamak to ITER. Such plasmas rely on a virtuous regime based on an interplay between density, particle fuelling, rotation and plasma composition which generates at the same time small ELMs, high core confinement and fusion reactions and reduced or even suppressed core W accumulation.

\emph{Experimental results}.$-$ Recent experiments in JET have demonstrated new H-mode operating regimes with simultaneous access to small ELMs and high thermal energy confinement, with high D-D fusion reactions and low core high Z impurity accumulation. Such regimes has been achieved using several techniques which rely on the optimization of the D neutral gas particle fuelling by either completely removing it or by partially replacing it with pellet injection \cite{Lang_2004}. Further addition of small Neon quantities has proven to be beneficial for ensuring a stationary plasma for several energy confinement times. 
In order to exemplify the type of plasmas obtained with small ELMs, two  discharges, \#94442 and \#96994 are compared to a typical type-I ELM discharge, \#97395, in Figure \ref{summary}. All these discharges share a common configuration, also called baseline, with toroidal current, $I_p$=3MA, toroidal magnetic field, $B_t$=2.8T, $q_{95}$=3.2 and low triangularity configuration, $\delta=0.2$. 3MW of Ion Cyclotron Heating (ICRH) were applied to all the discharges in addition to the Neutral Beam Injection (NBI) heating. The pulses \#97395 and \#96994 have the same total input power, 33 MW, however the particle source is provided in a very different way. Whereas for the pulse \#97395, the deuterium source is provided with D gas puff, with a total electron flux rate of $\Gamma_e(D)=2.8\times10^{22}s^{-1}$, in the case of \#96994 it has a mixture of D gas puff at a reduced rate, $\Gamma_e(D)=1.2\times10^{22}s^{-1}$ and 2mm D pacing pellets at frequency $f$=45Hz. Recent experiments showing that good pedestal and core thermal energy confinement can be obtained in JET with neon injection and small ELMs \cite{Giroud_2021}, motivated the addition of neon quantities to the baseline plasmas discussed here. The quantity of Ne injected, $\Gamma_e(Ne)=0.5\times10^{22}s^{-1}$, is small compared to the one used in highly radiating scenarios \cite{Gloggler_2019} and was limited by the fact that at high Ne injection, the neutron rate can significantly decrease \cite{Challis_2017}. Although the Greenwald fraction, $F_{gr}$, defined as the ratio between the average electron density to the Greenwald density $n_{GW}=10^{20}(m^{-3})I_p(MA)/\pi a^2(m^2)$ with $a$ the minor radius, is very similar for both discharges, $F_{gr}=0.7$, the differences in particle sources lead to a remarkably different ELM behaviour as shown in Figure \ref{summary}. In the case of \#97395, regular Type-I ELMs at $f_{ELM}$=45Hz are obtained, with relative averaged plasma energy change during individual ELMs of $\Delta W_{MHD}/W_{MHD}=2.2\%$. However, the discharge \#96994 has a compound ELM behaviour, characteristic of JET baseline plasmas with low gas puffing and pellets \cite{Garzotti_2019}, with isolated large ELMs followed by long periods of faster and smaller ELMs, with frequencies up to $f_{ELM}$=400Hz, resulting in a significant reduction of the averaged energy ELM losses, $\Delta W_{MHD}/W_{MHD}=0.8\%$, as shown in Figure \ref{summary}. In terms of performance, \#96994 performs better as both the thermal stored energy and neutron rate production are higher than the standard Type-I ELMy H-mode (\#97395). Furthermore, unlike what usually happens in Type-I ELMs regimes, which are prone to develop core W accumulation when the gas puff applied is low and therefore the $f_{ELM}$ obtained is low \cite{Angioni_2014}, in the discharge \#96994 the total core radiated power remains steady with no symptoms of impurity accumulation in the inner core. 
It is important to stress that, in terms of global parameters, the discharge \#96994 is close to what it is expected for the baseline scenario in ITER \cite{Sips_2018}, $H_{98}(y,2)=1.05$, $\beta_N=2.1$, $q_{95}=3.2$, $F_{gr}=0.7$ although the collisionality at the pedestal is slightly higher, $\nu^*_e$=0.4.

In order to evaluate the role of operating at low gas fuelling in the simultaneous access to the good confinement and small ELM in the discharge \#96994, a further reduction to zero gas injection has been performed in the discharge \#94442, heated with 17 MW of NBI power. In this case, D gas is injected in the initial phase of the discharge to control the entry to the H-mode, but it is completely switched off at t=9s, as shown in Figure \ref{summary} and no additional particle source is added. This leads to a phase with increased $f_{ELM}$ (up to 600 Hz) and very small ELM losses (with $\Delta W_{MHD}/W_{MHD}=0.6\%$ and $\Delta W_{MHD}<50 kJ$), during which the electron density continuously decreases, reaching a stationary value of $F_{gr} = 0.3$ at t=11.4s. At t=12s the gas puff is resumed and the discharge recovers its high density. During the phase where no gas in injected, the thermal energy confinement and neutron rate continuously increase, reaching values close to the discharge \#97395 which is heated with more NBI power. The radiated power does not show any transient behaviour and remains stable and comparable to the phase with gas injection. No impurity accumulation in the inner core is neither detected. Whereas the density obtained in this pulse is much lower than the ones expected in the ITER baseline scenario the pedestal collisionality is actually very similar, $\nu^*_e$=0.1. 

Such results indicate the key role of particle fuelling to control the transition to small ELMs plasmas with high confinement as for both discharges \#96994 and \#94442 the ELM behaviour looks quite similar regardless pellets are injected or not. However, the pellets provide enough fuelling in order to have a density compatible with high neutron rate production. These results are relevant for ITER as the main fuelling mechanism will be pellets rather than gas puff, which is expected to poorly penetrate the pedestal. 

\begin{figure*}
	\includegraphics[width=1\textwidth]{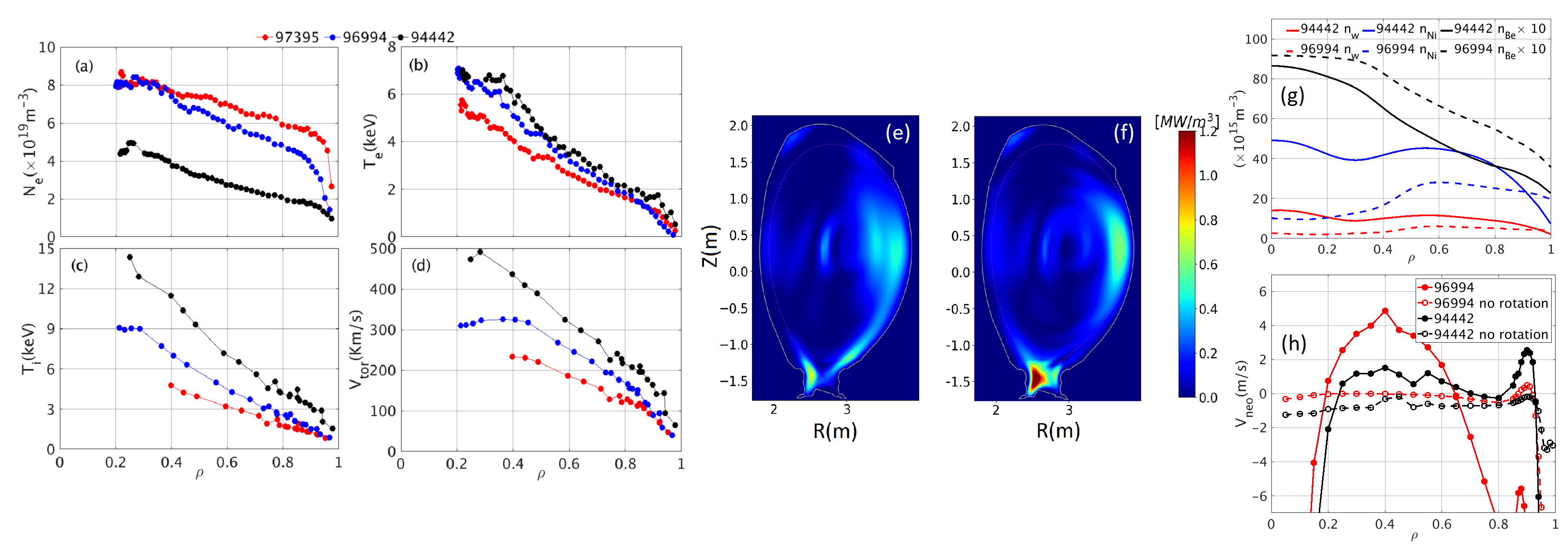}
	\caption{Electron density ($N_e$) (a), Electron temperature ($T_e$) (b),  Ion temperature ($T_i$) (c) and plasma toroidal rotation ($V_{tor}$) (d) for the discharges shown in Figure \ref{summary} at t=11.4 for \#94442, t=12s for \#97395 and t=12.5s for \#96994. Reconstructed bolometry radiated power 2D profile for the discharge \#94442 (e) and \#96994 (f). Tungsten ($n_W$), Nickel ($n_{Ni}$) and Beryllium ($n_{Be}$) density profiles for the discharges \#94442 and \#96994 (g). Convective neoclassical pinch velocity ($V_{neo}$) for the discharges \#94442 and \#96994 including and excluding the toroidal rotation in the NEO calculations (h)}
	\label{profiles}
\end{figure*}

\emph{Confinement analyses}.$-$ A detailed confinement study has been performed in order to clarify the physical mechanisms by which such plasmas with small ELMs regimes lead to good energy confinement. Clear differences in the electron and ion temperatures, electron density and rotation profiles are already evident as shown in Figure \ref{profiles}. Notably, the ratio $T_i/T_e$, including at the pedestal top, and the toroidal rotation significantly increase for the cases with small ELMs. This is accompanied by a reduction of the density at the top of the pedestal. In particular, for the discharges \#96994 and \#97395, at the same power and therefore torque, the central density is nearly identical, whereas the density at the pedestal is lower by $25\%$ and core rotation is $36\%$ higher at $\rho=0.4$ for the pulse \#96994, where $\rho$ is the normalized toroidal flux. Such features show that momentum confinement is also improved with small ELMs as it usually happens in JET plasmas with low gas fuelling and pellet injection \cite{Kim_2018}. This is a particularly important point for ITER as the rotation profile for the discharge \#97395 is close to what can be expected in ITER \cite{Chrystal_2020}.

Interestingly, the pedestal pressure remains nearly constant at, $P_{ped}=34kPa$ for \#96994 and $P_{ped}=36kPa$ for \#97395. Particularly striking are the differences in the pedestal structure. Both the electron density and temperature pedestals become wider and there is a substantial reduction in the edge density gradient as the edge density decreases. Compared to \#97395, where the pedestal top is located at $\rho=0.94$, for the discharges with small ELMs the pedestal top is located at $\rho=0.90$. Such differences in the pedestal structure lead to a decrease of the pressure gradient within the pedestal region for the discharges with small ELMs. Further analyses have shown that the pedestal in small ELMs plasmas are not limited by the Peeling-Ballooning modes which are characteristic or type-I ELMs \cite{Snyder_PB}.   
One key point of the small ELMs plasmas is the strong density peaking, which allows high core density at relatively low pedestal density. In such conditions high fusion neutron rate is obtained as it is generated in the inner plasma core. Unlike previous JET results in high thermal confinement plasmas with type-I ELMs, which also lead to high density peaking during the stationary phase \cite{Casson_2020}, the small ELMs plasmas analyzed in this letter do not show any kind of stronger radiation pattern in the inner core plasma, $\rho<0.3$. This is clearly shown in the 2D radiation profiles from bolometric tomography reconstruction shown in Figure \ref{profiles}. The impurity profiles, which are calculated as explained in \cite{sertoli_2019} and are consistent with data from the multiple diagnostics across the whole plasma radius, clearly demonstrate that the main impurity in the inner core is Be whereas high Z impurities such as Ni and W are either homogeneously distributed for the discharge \#94442 or even have a hollow profile for \#96994 with a maximum close to the pedestal as shown  in Figure \ref{profiles}. Maximum radiation at the pedestal region is common in plasmas with low gas puff and pellets, however in the case of \#96994, with the extrinsic impurity Ne injected, the radiation is redistributed towards a clear maximum at the divertor which is beneficial to avoid strong radiation in the plasma confinement region and yet, unlike in highly radiating scenarios with Ne \cite{Gloggler_2019}, the plasma remains attached to the divertor. 

Previous studies have shown that Tungsten is the main cause for the loss of plasma confinement and eventually radiation collapse in JET plasmas with high impurity accumulation. The transport of W from the periphery to the core plasma is mainly driven by collisional (neoclassical) convection which in conditions of high density peaking is usually strongly inward \cite{Angioni_2014}.    
W transport is analyzed in this Letter by studying the neoclassical transport, in particular the convective part of the flux with the gyrokinetic code NEO \cite{Belli_2008}. As shown for the discharges \#96994 and \#94442 in Figure \ref{profiles}, the W pinch is broadly positive (outward) for different plasma regions. Whereas for the discharge \#94442 the outward W pinch is high at the pedestal top and significantly positive in the whole region $0.2<\rho<0.9$, for the discharge \#96994 the outward W pinch is very strong in the region $0.2<\rho<0.65$, coinciding with the lack of radiation. Of particular importance is the role of rotation as it reverses the pinch from negative to positive in spite of the fact that plasmas 2D asymmetries by means of centrifugal forces tend to increase the inward W pinch \cite{Angioni_2014}. Therefore, for the first time at JET, it is shown that high confinement plasmas can be inherently resilient to strong inner W accumulation, as expected in ITER.   

\begin{figure}[hbt!]
\centering
	\includegraphics[width=0.48\textwidth]{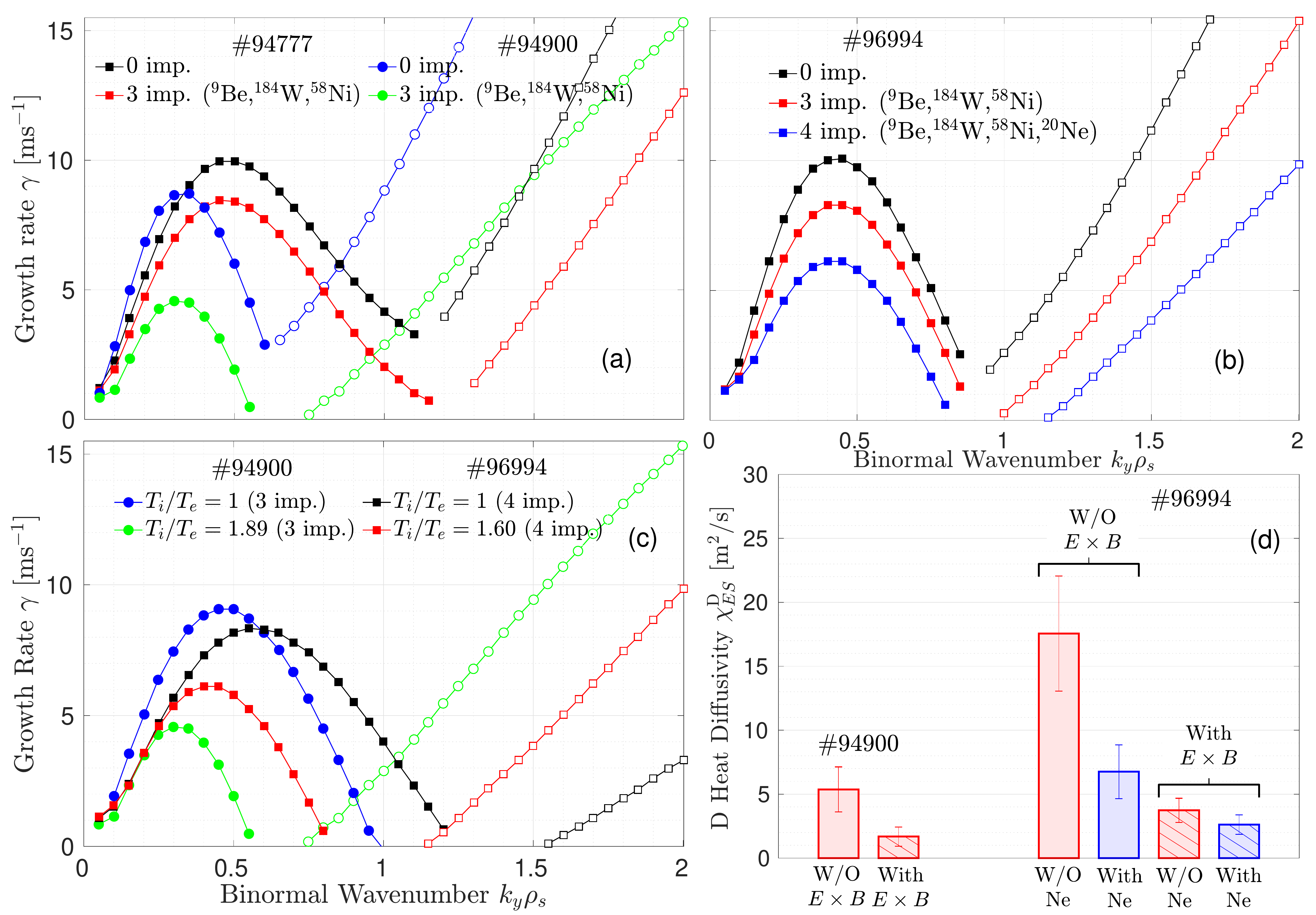}
	\caption{Turbulence growth rates obtained with the GENE code for the discharges \#94777, \#94900 and \#96994. Solid symbols represent ITG modes whereas empty symbols represent ETG modes (a,b) Growth rate comparison for the discharges \#94900 and \#96994 assuming experimental $T_i/T_e$ and artificially reduced to $T_i/T_e$=1 in both cases (c) D heat diffusivity obtained from non-linear simulations (d)}
	\label{turbulence}
\end{figure}

A detailed core transport analysis has been performed by means of GENE code \cite{jenko2000electron}, which solves the gyrokinetic equation, in order to evaluate the origin of the high confinement with the small ELMs. The analysis is focused at the radial location $\rho=0.8$ as it is found that the main differences between small ELMs and type-I ELMs are mainly localised close to the pedestal. For such analyses, the discharges \#94900 and \#94777, with Ip=3MA, Bt=2.8T, $q_{95}=3.2$ and low triangularity configuration are used as they are a match in all the parameters except on the gas puff used, removed for \#94900 (with small ELMs) and sustained for \#94777 (with type-I ELMs). For comparison, the pulse \#96994 is also analyzed. The results of linear simulations  are shown in Figure \ref{turbulence}. The maximum growth rate for the pulse \#94777, $\gamma_{max}^{94777}=10ms^{-1}$, is higher by 10\% than the one for \#94900, $\gamma_{max}^{94900}=9ms^{-1}$, whereas in both cases the linear spectra show that the dominant instability is the Ion Temperature Gradient (ITG) \cite{romanelli1989ion}, mainly driven by the steep thermal ion temperature gradient. Electron Temperature Gradient (ETG) modes \cite{Horton99}, are also identified at smaller spatial scales. The difference in growth rate is stronger when the impurities located at such radial location, Be, Ni and W are taken into account. For $\gamma_{max}^{94777}$ the reduction is very weak when the impurity concentration is included, however in the case of $\gamma_{max}^{94900}$ is reduced by 50\%. This shows that the edge located impurities, blocked by the neoclassical transport, have a beneficial effect on reducing transport driven by turbulence. In the case of the pulse \#96994, the trend is similar to the pulse \#94900, with $\gamma_{max}^{96994}=10 ms^{-1}$ which is also significantly reduced when including impurities as shown in Figure \ref{turbulence}. As found in highly radiating scenarios, the Ne injected also contributes to the ITG stabilization through dilution \cite{Gloggler_2019}. Additionally, a linear simulation has been performed for the pulses \#94900 and \#96994 by artificially decreasing the experimental value of $T_i/T_e$ to 1.00 as shown in Figure \ref{turbulence}. The effect on $\gamma_{max}$ is massive in both cases, with an increase of 50\% for \#94900 and 40\% for \#96994. This feature points out to the important role of the pedestal as a source of core improved confinement since, unlike for the pulse \#94777, for which $T_i/T_e=1.00$ at $\rho=0.9$, $T_i/T_e=1.69$ for the pulse \#94900 and $T_i/T_e=1.50$ for the pulse \#94900. 
Non-linear simulations have also been performed in order to analyze the impact of the ${E{\times}B}$ shearing, which is known to reduce heat transport by ITG \cite{doyle2007plasma} specially in conditions of low gas injected \cite{Kim_2018}. As shown in Figure \ref{turbulence}, the strong effect of ${E{\times}B}$ leads to a reduction of the deuterium diffusivities by three to four times. Such effect, in addition to the localized edge impurities (including Ne) and the high $T_i/T_e$ ratio starting at the pedestal, can explain the good confinement obtained in the small ELMs plasmas. 

\emph{Discussion and conclusions}.$-$ The results discussed in this Letter show that an optimum self-organized plasma state is obtained at JET, in the presence of a metallic wall, by controlling the particle fuelling conditions. The reduction and tailoring of the injected gas puff leads to the onset of small ELMs which in turn trigger hollow impurity profiles and increased rotation and its shear. In such conditions, edge turbulent transport is reduced and the direction of the impurity convection is outward. Whereas the highest performance is obtained when the gas puff is fully removed and the electron density strongly drops, stationary ITER relevant plasmas are only obtained when the gas puff is partially replaced by pacing D pellets. In such conditions, and compared to type-I ELMs baseline plasmas, the ion temperature and the toroidal rotation significantly increases in the whole plasma radius, the pedestal density drops and the inner core density remains unchanged. This is particularly important for ITER, for which extrapolated rotation values based on type-I ELMs plasmas fuelled with gas puff maybe be too pessimistic \cite{Chrystal_2020}. The fact that the neutron rate increases during the small ELMs regime indicates that such conditions are adequate for fusion power generation in metallic wall devices. Therefore, these results open up the possibility that magnetically confined plasmas self-organize in the direction required for optimum fusion power generation meanwhile strong heat fluxes to the plasma facing components are avoided.

\section*{Acknowledgments}
J. Garcia would like to thank Xavier Garbet and  Gerardo Giruzzi for fruitful discussions.

The gyrokinetic simulations were performed on CINECA Marconi HPC within the project WPJET1.

This work has been carried out within the framework of the EUROfusion Consortium and has received funding from the Euratom research and training programme 2014–2018 and 2019–2020 under Grant agreement No 633053. The views and opinions express herein do not necessarily reflect those of the European Commission. This research was supported in part by grant FIS2017-85252-R of the Spanish Research Agency, including ERDF-European Union funding.



\bibliography{MyBiblio}

\end{document}